\documentclass[12pt,preprint]{aastex}
\begin{document}

\shorttitle{Modeling Black Hole Retention}
\shortauthors{Moody \& Sigurdsson}

\title{Modeling the Retention Probability of Black Holes in Globular Clusters:  Kicks and Rates}

\author{Kenneth Moody, Steinn Sigurdsson}

\affil{Dept.\ of Astronomy and Astrophysics,
The Pennsylvania State University,
525 Davey Lab. University Park,PA 16802}

\email{<moody,steinn@astro.psu.edu>}

\begin{abstract}
We simulate black hole binary interactions to examine the probability of mergers and black hole growth and gravitational radiation signals using a specific initial distribution of masses for black holes in globular clusters and a simple semi-analytic formalism for dynamical interactions.  We include 3-body recoil and the latest results in numerical relativity for gravitational radiation recoil.  It is found that while 99\% of binaries are ejected from low metallicity, low mass clusters; metal rich massive clusters retain 5\% of their binaries.  An interesting fraction of the ejected binaries, especially those from high mass, high metallicity systems, merge on timescales short enough to be gravitational radiation sources during their mergers with rates approaching those expected for galactic field black hole binaries.  While the merger rates are comparable, the much larger mass of these binaries and their localization will make them appealing targets for advanced LIGO.  We single out two possible Milky Way clusters (NGC 6441 and NGC 6388) as having the properties for a good probability of retention.
\end{abstract}


\section{Introduction}

Observed black hole masses occupy two regimes, $M_{BH}\lesssim100M_{\sun}$ for black holes formed from core collapse supernova, and supermassive black holes with $M_{BH}\gtrsim10^6M_{\sun}$ which reside in the centers of galaxies.  Observations of some objects, however have suggested that a middle regime of intermediate mass black holes (IMBH, review by \citet{Miller04}) could exist with masses between stellar and supermassive black holes.  Ultraluminous X-ray sources ($L_X>10^{39}$ ergs sec$^{-1}$) have been found in intense star forming regions outside the nuclei of some galaxies \citep{Kaaret01,Matsumoto01,Fabbiano01}, and recently even in one globular cluster \citep{Maccarone07}.  The lower limit on the mass for these objects, assuming isotropic emission at the Eddington limit, is a few hundred solar masses.  A stellar mass black hole would require special geometry of its accretion disk for sufficient beaming to occur and the accretion to be sub-Eddington \citep{King01}, or need special conditions on the gas to provide a super-Eddington accretion rate \citep{Begelman01}.  Conversely, a supermassive black hole ($M\gtrsim10^6 M_{\sun}$) would experience dynamical friction and sink to the center of its galaxy in too short a time to be plausibly observed at locations within the host galaxies where the IMBH candidates are projected to be seen today \citep{Kaaret01}.  An intermediate mass for these objects would seem to be indicated.  In the nearly bulgeless galaxy NGC4395, there has been found \citep{Fillipenko03} an AGN the black hole mass for which seems to be $\lesssim 10^5M_{\sun}$, which is in the upper range of IMBH masses.  An object with a similar mass may also exist in the galaxy POX 52 \citep{Barth04}.

Another possible place to look for IMBHs besides in starburst regions in galaxies is in the centers of globular clusters.  Observations have seen an increase in the mass-to-light ratio towards the centers of two globular clusters that might be consistent with a massive object, a $M>10^4M_{\sun}$ object in the Andromeda Galaxy cluster G1 \citep{GRH02, GRH05}, a $4\times10^4M_{\sun}$ object found by \citet{Noyola08} in the cluster $\omega$ Cen, and a few thousand solar mass object in M15 \citet{vanderMarel02, Gerssen02}, although in the case of M15, \citet{Baumgardt03} are able to simulate the observations without an IMBH using smaller compact objects.  The velocity dispersion of the central stars in the cores of these globular clusters as compared to the conjectured mass of the IMBH put these clusters on the same M-$\sigma$ relation as the bulges of galaxies with supermassive black holes \citep{GRH02, vanderMarel02, Gerssen02}.  While the theory of the origin of the M-$\sigma$ relation for supermassive black holes would probably not apply to globular clusters, it is intriguing that, at least in these two cases, the IMBHs in these clusters are consistent with it.
 
Two possible formation scenarios for the formation of IMBHs within a globular cluster have been proposed recently.  The first involves the process of core collapse, by which the heavier stars in a cluster first sink to the middle through mass segregation.  The stellar density goes very high, and might be sufficient that several stars collide forming a very large star (mass a few hundred solar masses), which collapses directly to an IMBH \citep{Begelman78,PortZ02,Freitag07}.  Otherwise, stellar evolution causes the high mass stars to form black holes, which can become binaries through exchange into existing binaries of lower mass main sequence or neutron stars \citep{Sigurdsson93b}.  We assume that formation of an IMBH by runaway merger does not occur in this case.  Three-body interactions, which in the cores of clusters are dominated by interactions with all three objects being black holes, can then begin to work to harden binaries to the point where they merge.  It is this scenario we intend to investigate.  

Previous studies of dynamical formation of a IMBH from stellar mass black holes have been performed.  Black holes have been shown to be dynamically important in such aspects as the radius-age relation of Magellanic Cloud clusters \citep{Mackey07,Mackey08}, and even in galactic nuclei \citep{Lee95}.  \citet{PortZ00} include a study of how important black hole binaries in clusters are to gravitational wave research.  Recently, \citet{HB07} did a study of mergers of black holes in a system already containing a few hundred solar mass black hole, which represents the next step in IMBH formation after our work.  \citet{Kulkarni93} and \citet{Sigurdsson93a} used only 10 M$_{\sun}$ black holes, and determined that the formation of 10$^3$ M$_{\sun}$ objects is possible.  On the other hand, \citet{Miller02} showed that these binaries tend to be ejected before reaching a size at which recoil becomes unimportant, precluding further growth.  Since black holes are produced from progenitors with a wide range of masses (20-100 M$_{\sun}$) and have varied evolution histories (wind losses and mass transfer) just prior to becoming black holes, a distribution in masses may better reflect the actual situation in globular clusters.  \citet{OLeary06} did a study using the distribution of black hole masses and binary periods as given in \citet{Belczynski04} and a more complicated method of computing interactions than the semi-analytic model we use.  They use the old prescription for gravitational radiation recoil similar to that found in \citet{Favata04}.  Numerically simulating black hole mergers through ringdown has now been done \citep{Gonzalez06} and definitive recoil velocities determined, so that the sole remaining uncertainty in determining the circumstances of black hole mergers is the initial distributions of their masses.  Our semi-analytic method can more quickly respond to updates in stellar population synthesis models than direct many-body integration.  

We describe the conditions under which the simulations were done, including initial conditions of the binary and the analytical form of the 3-body interactions and relevant time scales, then report results from several simulation runs, with a few parameters (e.g. metallicity) adjusted after each one.  Finally, we discuss what the results imply for observed systems and suggest two systems that fall in the higher probability category for harboring an IMBH.

\section{Simulation Conditions}

We simulate the history for a total of 100,000 binaries for each set of initial conditions.  The ensemble of initial conditions includes two values each for metallicity and escape velocity.  An examination of \citet{Belczynski04} shows that there exist only two distinct shapes of the period and mass distributions based on metallicity, therefore we only include a qualitative distinction with the changeover coming at $\log [Fe/H]=-1.3$ for observed clusters.  The escape velocities were chosen as a proxy for several properties to represent moderately sized clusters and heavy clusters; smaller clusters which might have had a lower value for the escape velocity are not expected to retain any binaries that interact due to their extremely shallow potentials.  The models' initial conditions are described in Table~\ref{modeltab}.  Each binary history is run until one of three fates is determined:  ejection as a binary through 3-body superelastic recoil (Fate 1), ejection as a single object upon merger from gravitational radiation recoil (Fate 2), or a retained merged single object (Fate 3).  For those systems which come under fate 1, we also calculate the gravitational radiation merger timescale for the ejected binary and determine the fraction of those in the run that coalesce within a Hubble time.  Results for the runs are in Table~\ref{restab}.

Whereas previously (c.f. \citet{Sigurdsson93a}) the distribution of black hole masses has been assumed to be single-valued at 10M$_{\sun}$, we start with a multi-valued initial mass function (IMF) for black holes.  Based on figure 3 of \citet{Fryer01} for f=1 (fully efficient supernova) and stellar IMF power-law index $\gamma$=2.0, we propose using for our low metallicity distribution a smooth power law, which gives the probability of a black hole having mass $M$ proportional to $10^{-.05M/M_{\sun}}$ for masses in the range 3M$_{\sun}$ to 80M$_{\sun}$.  This is the same analytic form as the low metallicity (Z=0.001) IMF found by \citet{Belczynski04} for his standard model parameters.  \citet{Fryer01} calculate a black hole IMF for single progenitor stars, and are used by \citet{Belczynski04} to predict black hole masses from a range of progenitor masses, while also considering binary evolution effects such as common envelope phases.  The most uncertainty in determining mass functions from binary evolution models comes during the common envelope phase and in calculating mass loss.  These are common to any attempt to use a distribution of masses.  

The IMFs in \citet{Belczynski04} are presented as histograms, but we convert these to analytical probability distributions for ease of use with computer-based simulations.  Binaries are constructed using the IMF to pick both masses, a distribution in periods taken from \citet{Belczynski04} as appropriate for the metallicity studied and also converted to an analytical form, and an eccentricity from a thermal distribution (P(e)=2e).  We chose the masses independently as we expect the stars to have, for the most part, developed independently in well separated binaries.  The simulations take place in a regime, being the center of a dense cluster, where all of the stars not heavy enough to have become evolved have been ejected to the outskirts of the cluster, except in the first ~100Myr when massive stars were evolving rapidly and drastically changing core conditions. We assume that we start after this time, unlike other studies such as \citet{PortZ02}, and have a quasi-static cluster environment.  The forms for both the IMF and periods are as follows \citep{Belczynski04}:
$$P_{{\rm low Z}}(M)=0.152\cdot10^{-0.05M/M_{\sun}}{,\,\,\rm all\,\,M}.$$

$$P_{{\rm high Z}}(M)=\cases{0.028, &$3M_{\sun}\geq M\geq 15M_{\sun}$, \cr 10^{0.6M/M_{\sun}}, &$15M_{\sun}\geq M\geq 55M_{\sun}$, \cr 0, &$55M_{\sun}\geq M\geq 80M_{\sun}$.}$$

$$P(P)=\cases{\sqrt{1-(P-1)^2}, &$0\geq \log P({\rm days})\geq 2$ low Z only, 
\cr (P-2), &$2\geq \log P({\rm days})\geq 6$ all Z.}$$

Binaries are subjected to encounters with a third black hole whose mass is randomly drawn from the IMF.  The time scale for the encounter $t_{enc}$

\begin{equation}
t_{enc}=1.5\times10^9 {\rm years} \frac{m_3 v_{10}}{\mu_{1\,2}m_T a_{AU}n_4}
\end{equation}

\noindent
is calculated from equation (2.9) of \citet{Sigurdsson93b} with $m_3$ the mass of the third black hole, $\mu_{1\,2}$ the reduced mass of the binary, $m_T$ the total mass of all three objects, the masses being expressed in terms of $M_{\sun}$, $a_{AU}$ the binary's semimajor axis, $n_4$ the density of stars $n/10^4$ pc$^{-3}$, and $v_{10}$ the relative velocity of the third object $v/10$ km/s, which is taken to be 1 for typical globular clusters, which have velocity dispersions on the order of 10 km/sec.  For this work we use a value of the dimensionless cross section $\tilde{\sigma}=10$ as defined in equation (2.7), as this value is broadly consistent with interacting systems having mass ratios in the range of those in our simulations as given in tables 3A and 3B of \citet{Sigurdsson93b}.  This time scale is compared to the time for merger by emission of gravitational radiation \citep{Peters64}

\begin{equation}
t_{GW}=3.151\times10^{17}{\rm years}\ g(e) \left(\frac{a}{AU} \right)^4 \left(\frac{M_{\sun}}{m_1}\right) \left(\frac{M_{\sun}} {m_2}\right) \left(\frac{M_{\sun}}{m_1+m_2}\right)
\end{equation}
\begin{equation}
g(e)=\left(1-e^2\right)^{7/2} \left(1 + \frac{73}{24}e^2 + \frac{37}{96}e^4 \right) 
\end{equation}

For $t_{enc}<t_{GW}$, the encounter takes place.  This entails choosing a new eccentricity from the thermal distribution and a change in the semimajor axis $a$ such that the binding energy of the binary is changed by 

\begin{equation}
\Delta = 1-\frac{a_{in}}{a_{out}}\frac{m_am_b}{m_1m_2}
\end{equation}

\noindent
where $a_{in}$ and $a_{out}$ are the starting and ending semimajor axes of the binary for the encounter, $m_1$ and $m_2$ are the masses of the two original objects in the binary, and $m_a$ and $m_b$ are the masses of the two new objects.  Based on the results in \citet{Sigurdsson93b}, we assume that the two most massive of the three interacting objects form the new binary, leading to the possibility of membership change.  If there is not a change, $a$ is simply reduced by a factor $(1-\Delta)$.  With an exchange in membership, it is possible for $a$ to increase dramatically. For simplicity, we choose a fixed value of $\Delta = 0.4$, characteristic of the mean energy transferred in the same encounters from \citet{Sigurdsson93b} that gave us our value for $\tilde{\sigma}$. This is warranted, if there are multiple encounters per system before ejection.  For a check on the adequacy of a single value of $\Delta$, we ran one set of simulations allowing $\Delta$ for each interaction to vary in a normal distribution around 0.4.  Even with a variance of 0.2, the effect was negligible.  We do not track stellar interactions, implicitly we assume that we are in a regime where there are multiple black holes which have formed a dense sub-core in the cluster, and that the interactions of these black holes dominates the fate of any binary. In general, interactions of the sub-core with stars are not important.  They become very important in late stages, particularly for the ''last'' black hole binary.  Discussion of stellar interactions with the binary is beynd the scope of this paper.

Besides changes in the internal dynamics of the binary, the conservation of momentum among the two systems requires the binary to recoil.  The magnitude of the recoil is

\begin{equation}
v_{rec}=\frac{m_e}{m_T}\sqrt{\frac{m_3(m_1+m_2)} {m_e(m_a+m_b)}\sigma_{GC} + \frac{2\Delta m_TGm_1m_2}{m_e(m_a+m_b)a_{in}}}
\end{equation}

If the recoil is smaller than the assumed escape velocity of the globular cluster, the time is incremented by $t_{enc}$, and the run continues by choosing a black hole mass independently from the mass distribution.  The dynamical friction timescale for the binary is approximately $\langle m\rangle/M_{BH} t_r$ \citep{OLeary06}, where $\langle m\rangle$ is the average stellar mass, and $t_r$ is the relaxation timescale.  Very few binaries are kicked with the narrow range of velocity required to have a turning point of several half-mass radii, and these are highly radial orbits.  Therefore, the core relaxation time applies.  The core relaxation times for most clusters are $10^7-10^8$ years \citep{Harris96}, so most binaries will return to the core in less than a million years.  If the binary is ejected, the run is stopped and $t_{GW}$ is calculated for the binary.  Those for which $t_{GW}$ is less than 10$^{10}$ years may be field gravitational radiation sources.  The run will also be stopped once $t_{enc}>t_{GW}$, at which time the recoil velocity from asymmetric emission of gravitational radiation is calculated.  We used the zero spin expression for the gravitational radiation recoil from \citet{Gonzalez06}

\begin{equation}
v_{GW}=1.20\times 10^4\eta^2\sqrt{1-4\eta}(1-0.93\eta)\ {\rm km/sec}
\end{equation}

\noindent
where $\eta$ is the symmetric mass ratio defined using $\eta=q/(1+q)^2$, $q$ being the mass ratio of the two objects in the binary ($0\geq q\geq1,0\geq \eta \geq \ 0.25$).  Stellar mass black holes, unlike supermassive black holes in galaxy centers which have accreted most of their mass from a thin disk where the spin goes to 0.98, are not likely to have a large spin parameter.  \citet{OShaun05b} find by analogy with neutron star birth spins \citep{Lorimer05,Kramer03,Migliazzo02} that expected spins should be less than 0.1, unless otherwise spun up by fallback from the supernova explosion.  \citet{Burrows07} find that only rapidly spinning cores may produce the phenomena called hypernovae.  Since these types of objects are rare, we may infer that most supernovae that produce black holes make slowly spinning ones.  Any accretion that does occur while the black hole is in a binary with a mass donor is expected by \citet{Belczynski08} to increase the spin parameter $a=J/M^2$ beyond 0.5.  From the fully spin dependent form of the recoil velocity \citep{Campanelli07} and using the approximation of $1/\sqrt{2}$ for the values of the sine and cosine of the angles, we find that $v_{GW}(a)/v_{GW}(a=0)$ goes above 2 for values of $a>\sim 0.4$ except for extreme mass ratios which are more sensitive to spin.  The merged object is then ejected or retained in the globular cluster depending on the magnitude of $v_{GW}$.

\section{Results}

\begin{figure}[ph]
\centerline{
\rotatebox{0}{\scalebox{0.45}{\includegraphics{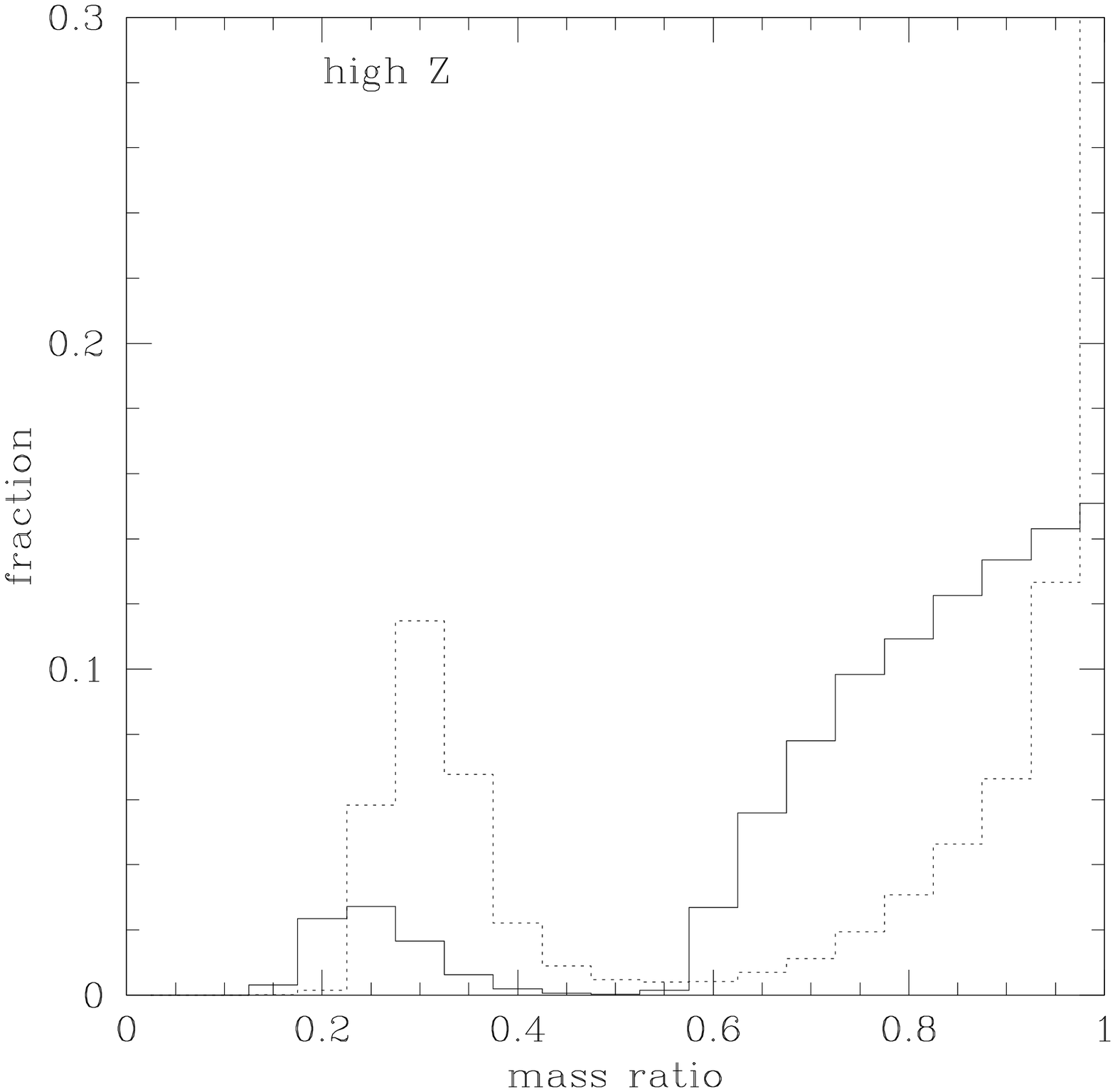}}}
} \centerline{
\rotatebox{0}{\scalebox{0.45}{\includegraphics{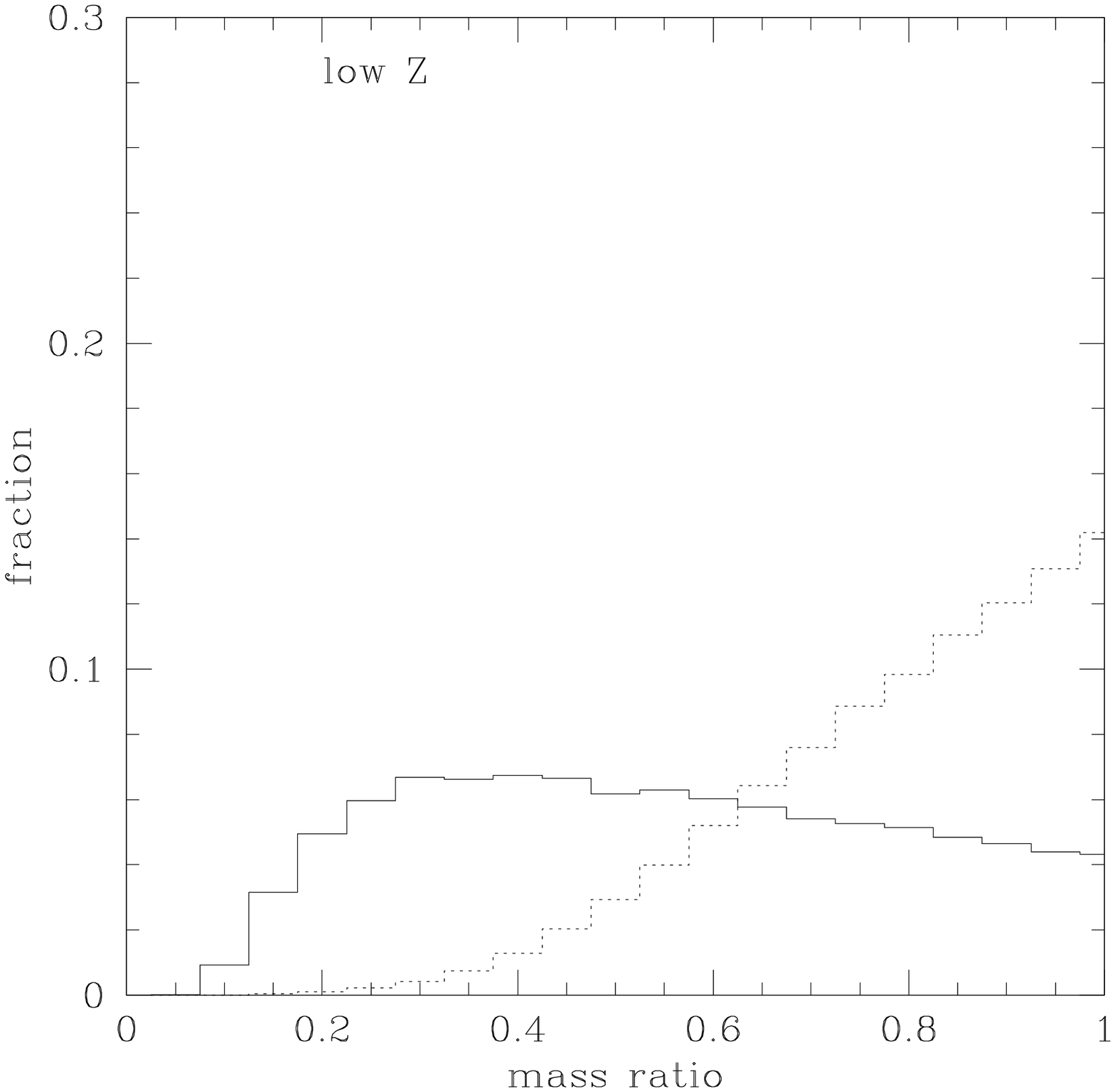}}}
}
\caption{Histogram showing distributions of mass ratios.  The top panel shows the metal rich initial condition, the bottom panel has the metal poor condition.  The solid line in each plot is the distribution obtained by randomly selecting two black holes from the indicated distribution, as is done initially in each simulation.  The dotted line shows the distribution of mass ratios for ejected binaries.  The initial high metallicity mass ratio distribution is bimodal due to the black hole IMF having two peaks near 10 and 45 M$_{\sun}$, giving a $q$ of ~0.25.  Three body recoil is efficient at ejecting binaries with mass ratios $q\gtrsim0.3$, therefore the distribution of the mass ratios of ejected binaries is skewed closer to 1.  In the high metallicity case, this includes the lower q peak shifting to 0.3\label{ratio}}
\end{figure}

As seen in Table \ref{restab}, the most likely conditions for a black hole to be retained are in massive, metal-rich clusters.  The change in mass distribution with metallicity is the main driver of whether or not a binary may be retained.  What changes most between the two metallicities is the distribution of mass ratios, seen in Figure~\ref{ratio}.  The initial distribution of mass ratios for metal poor binaries is nearly constant above 0.2, which is the condition \citet{OLeary06} place on their binaries {\it a priori}.  For the distribution of initial mass ratios for high metallicity binaries, there is a peak at $q=0.25$ due to systems with one member from each of the two parts of the distribution (centered at 10 and 40M$_{\sun}$).  The mass ratio distribution of ejected binaries is shifted toward higher values of $q$ for both high and low metallicity distributions, meaning more equal mass binaries are more likely to be ejected.  This follows from previous attempts at this problem \citep{Kulkarni93,Sigurdsson93b} with the failure of equal mass binaries to produce a retained object.  

We compare our results of ejected binaries to those of \citet{OLeary06} by collecting enough of our runs within a single model to create an N=512 cluster.  The models of theirs most similar to our were the e5e5king11, v2e5k11, and v3e5k11 models for low escape velocity, and the e5e5king7, v2e5k7, and v3e5k7 models for high escape velocity.  All of these fall into what we consider low metallicity models, as that corresponds to the models of \citet{Belczynski04} the authors used.  In the models indicated, \citet{OLeary06} find the ejection fraction of black holes in binaries in low $v_{esc}$ clusters is 0.14, and a fraction of 0.1 in high $v_{esc}$ clusters.  We find similar values for the fraction of ejected black holes in binaries.

\begin{figure}[ph]
\centerline{
\rotatebox{0}{\scalebox{0.6}{\includegraphics{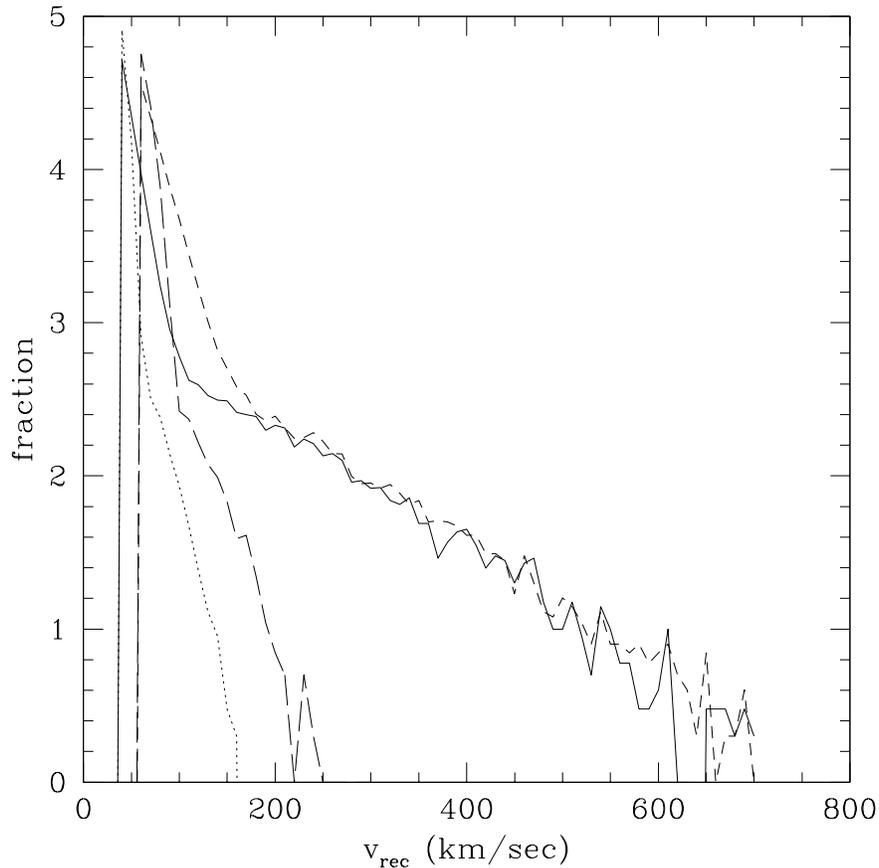}}}
}
\caption{Plot of the recoil velocities of ejected binaries.  The solid line is for model A (low mass, low metallicity).  The dotted line shows model B (low mass, high metallicity).  The short dashed line is for model C (high mass, low metallicity).  The long dashed line is for model D (high mass, high metallicity).  The plots start at the escape velocity for the cluster.  The high velocity tail for models A and C is due to binaries from the low metallicity period distribution which have small initial separations since $v_{rec}$ is inversely proportional to $a$.  This also is the reason why the high mass cluster models are shifted to higher velocities (at least for $v_{rec}$ between 50 and 200 km/sec), as the binaries are able to become more tightly bound before being ejected.  The fraction of binaries at each point covers a 5 km/sec bin. \label{vrec}}
\end{figure}

The large numbers of ejected binaries produced in each cluster make them an interesting target of investigation.  Some fraction of these we find have a $t_{GW}$ less than 10Gyr, and so would make a background source for gravitational radiation detectors \citep{PortZ00}.   Which property of the cluster is more important in determining the efficiency of producing binaries that will merge in less than a Hubble time is complex.  High mass metal rich clusters produce the most binaries that merge in less than 10Gyr.  The next most come from high mass metal poor clusters.  Low mass metal poor clusters produce more binaries that merge in less than 10Gyr than low mass metal rich clusters.  The change in Hubble time mergers with mass is expected as a heavier cluster would allow the binary to become harder before ejecting it.  The binaries for the most part stay within their host galaxy.  Figure~\ref{vrec} shows the distribution of velocities of the ejected binaries.  For the two low metallicity models, which have binaries that start with tight orbits having $\log P(days)<2$, there are a few systems (~1\%) that are ejected with a velocity higher than 300 km/sec, but most binaries (and all of the high metallicity ones) have $v_{rec}< 200$km/sec.  This means that while they leave their parent cluster, they are still confined to their parent galaxy unless it is a dwarf galaxy.

\begin{figure}[ph]
\centerline{
\rotatebox{0}{\scalebox{0.4}{\includegraphics{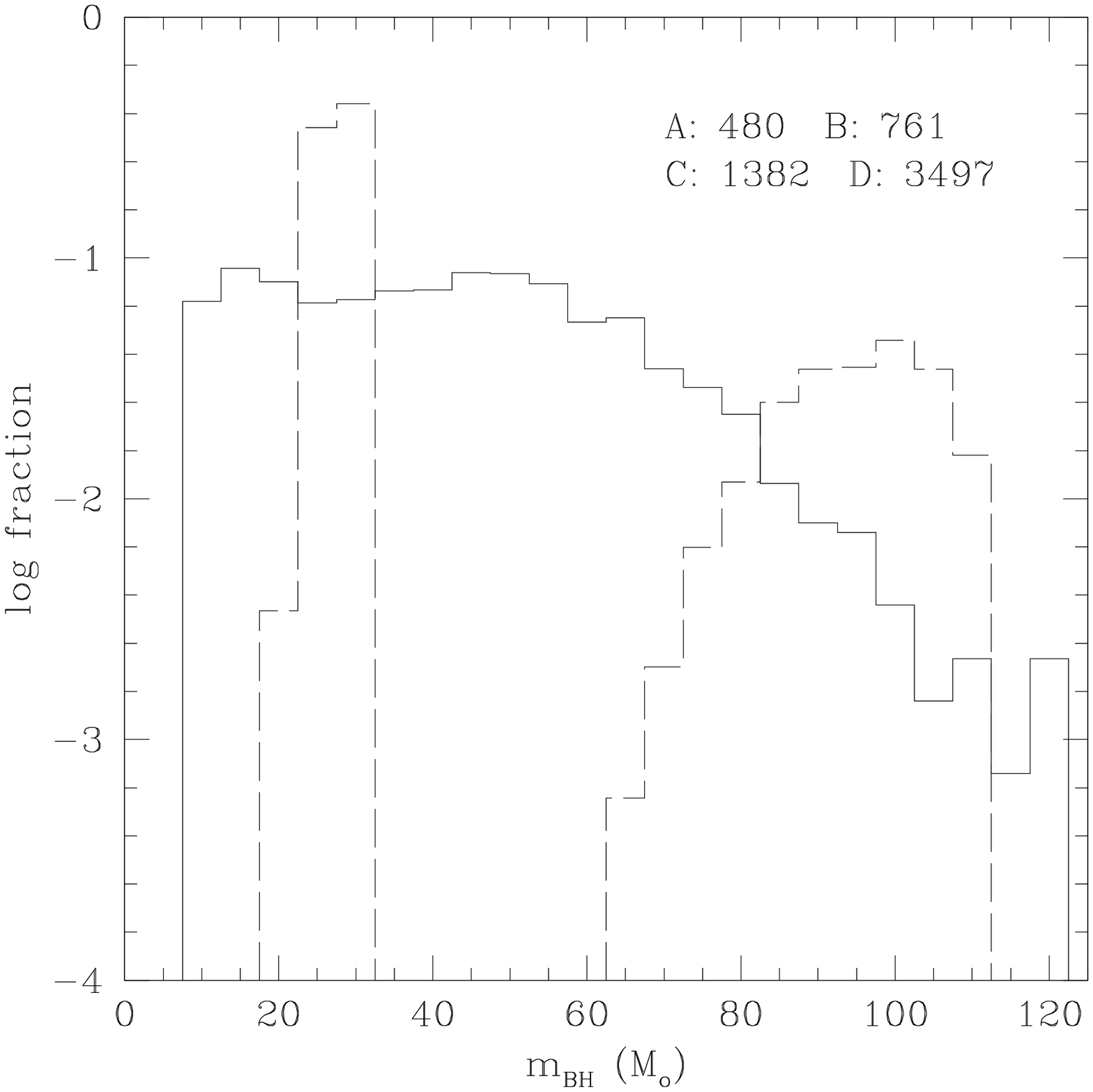}}}
\rotatebox{0}{\scalebox{0.4}{\includegraphics{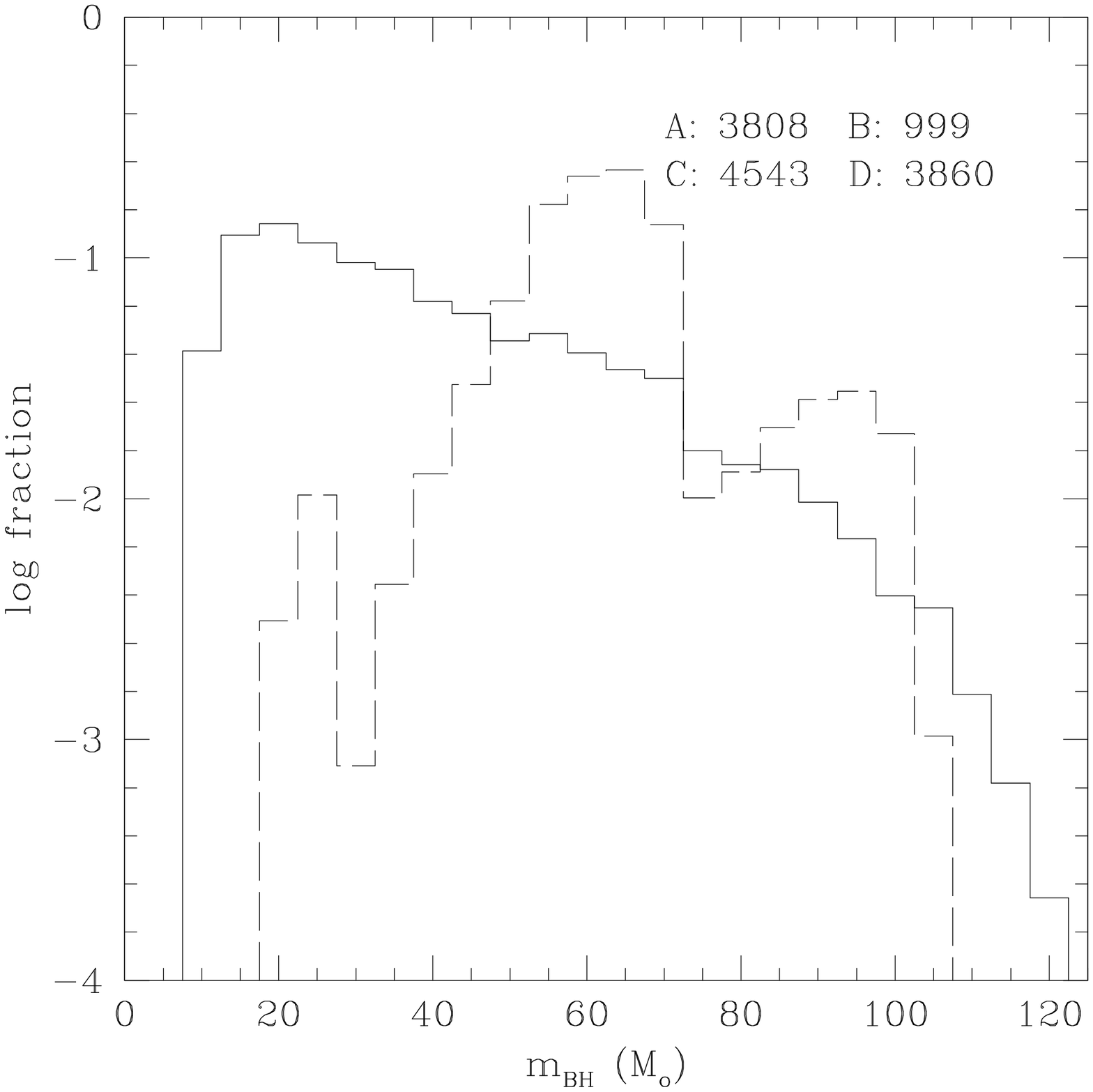}}}
}
\caption{The top histogram showing distribution of masses for the black holes retained upon merger.  The bins are 5M$_{\sun}$ wide and show the log of the fraction in each bin.  The solid line is for metal poor systems, while the dashed line is for metal rich systems.  The model dependence is most visible in the second of these with the lack of merged black holes at 40 to 60M$_{\sun}$ and the sharp dropoff above 110M$_{\sun}$.  We see that there are a substantial fraction (1\% to 5\%) of black holes that remain which have masses above 100M$_{\sun}$, which is a common definition for the lower boundary IMBH masses.  The bottom plot shows the mass distribution for those objects ejected upon merger by gravitational radiation, with the same convention for the lines.  Both of these plots are normalized to the number of black holes that undergo fate 2 (ejection) or fate 3 (retention), the number of which for each model is given.  These objects show a "complimentary" distribution to the retained objects, especially for the metal rich clusters where the peak mass of ejected objects fits nicely into the deficit of retained objects.\label{mass}}
\end{figure}

The distributions of masses for the retained merged objects are in Figure~\ref{mass}.  For low metallicity systems, the distribution is flat up to 60M$_{\sun}$, after which it drops.  High metallicity systems show one peak of 20-30M$_{\sun}$, and another between 80 and 100M$_{\sun}$, which reflects the underlying initial mass distribution.  For ejected objects, the low metallicity systems show a monotonic decline from 15M$_{\sun}$ to the maximum mass seen at ~120M$_{\sun}$, with a slight break downward at 60M$_{\sun}$.  For the high metallicity systems, there is a peak around 40-60M$_{\sun}$ where the distribution of masses for retained objects has a deficit.  The currently used recoil velocity function has its peak at a mass ratio of about a third, so that if a binary in the high metallicity model consists of one member from each of the two regions, it will have a total mass of about 40-60M$_{\sun}$ and a mass ratio of 0.3-0.4 which will most likely be ejected, whereas a binary with both members from the same region will have a mass of either 20 or 80-100M$_{\sun}$ and $q$ close to one and be retained (if it survives 3-body interactions of course).  

\begin{figure}[ph]
\centerline{
\rotatebox{0}{\scalebox{0.45}{\includegraphics{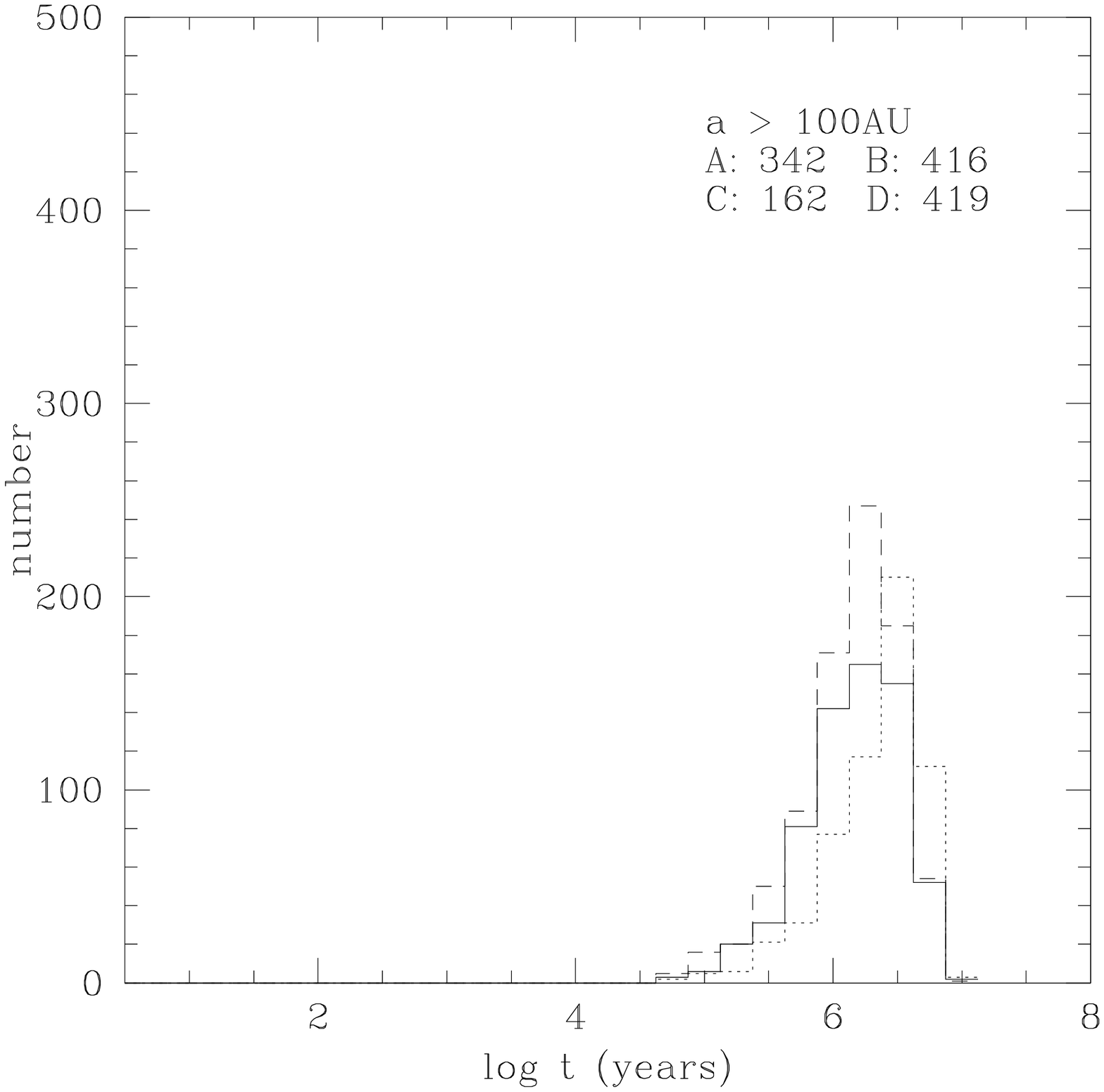}}}
\rotatebox{0}{\scalebox{0.45}{\includegraphics{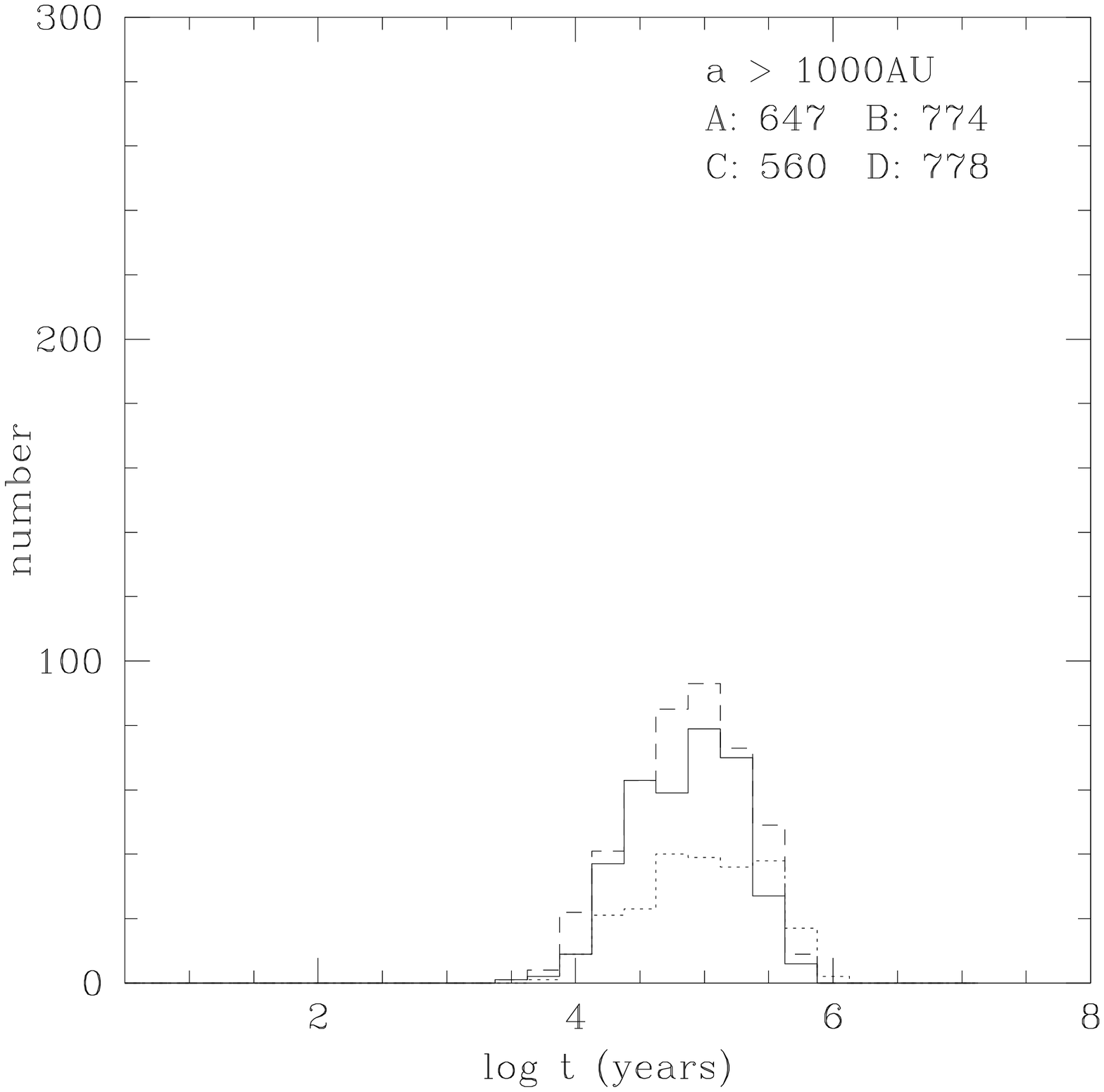}}}
} 
\centerline{
\rotatebox{0}{\scalebox{0.45}{\includegraphics{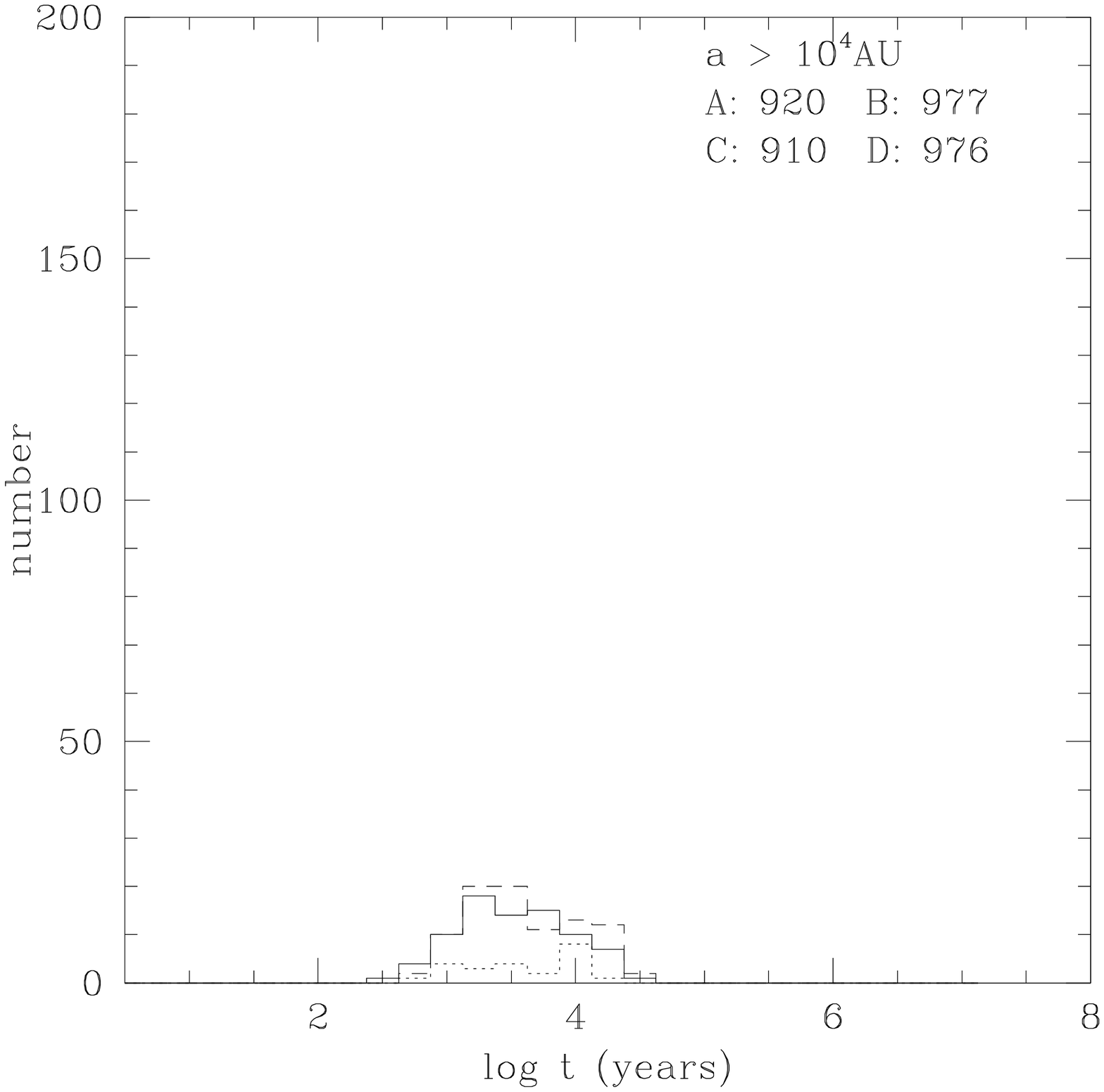}}}
}
\caption{Distributions of time spent at large $a$ by the 1000 binaries for which we kept detailed histories.  The solid line is for model A, the dotted for model B (which is identical to the distributions for model D), and the dashed line for model C.  The top left plot show the number of binaries that exist at an $a>10^2$ in 0.25 dex bins, the top right shows the same for $a>10^3$, and the bottom plot shows those which had $a>10^4$.  Numbers in the plot show the number of binaries for each model which never have the semimajor axis indicated.  The time spent by the binaries at semimajor axes of $a\gtrsim10^{3.5}$ is small due to 3-body interactions hardening the binary.\label{hia}}
\end{figure}

While the binaries are in the globular cluster, they may go through short-lived phases with large semi-major axes due to exchanges of membership.  These stages may be important in transferring angular momentum from the binary to the cluster as a whole through interactions with stars as shown in \citet{Mapelli05}.  For a fraction of the binaries, the histories of $a$ are recorded and examined to determine how much time they spend with $a>10^2$, $10^3$, and $10^4$ AU.  The distributions of the two high metallicity samples are the same, but the low mass metal poor model shows fewer binaries that get to high separations.  The low metallicity low and high mass models have 34.2\% and 16.2\% that never have $a>10$AU respectively, while for high metallicity this percentage is 41.6\%.  Other than this difference, the distributions of time spent at high separations is similar.  The plots for the number of binaries that exist at high separations for the time indicated are in figure~\ref{hia}.  

\section{Discussion}

The event rate from merging black hole binaries can be calculated from the fraction of systems that merge within a Hubble time and the relative contributions from low and high metallicity systems and light or massive clusters.  To conservatively estimate the event rate, we assume 100 globular clusters per galaxy (e.g. the Milky Way is currently thought to have about 150) and 100 BH per globular cluster ($N_{BH}\sim10^{-4}N_{\star}$).  We assume that the break between high versus low metallicity is at an [Fe/H] of -1.3 and that light globulars have $v_{esc}$ of less than 30 km sec$^{-1}$ and heavy globulars above this value.  The escape velocity for a globular cluster is given by $v_{esc}=\sqrt{2\Phi_0}$, where $\Phi_0$ is the central potential of the cluster, $W=\Phi_0/\sigma^2$ is the King parameter and is correlated with the cluster concentration, and $\sigma$ is approximately equal to the velocity dispersion except in the case for shallow globulars.  We determine the concentration and thus $W$ from the catalog of \citet{Harris96}, while 1D velocity dispersion data were obtained from \citet{Pryor93}.  For the 56 Milky Way clusters for which we could determine the escape velocity, we find that the percentage of clusters in each of our models is as follows:  A 45\% (25), B 21\% (12), C 20\% (11), and D 14\% (8).  Including data from table~\ref{restab} on the number of mergers within a Hubble time, we find that ejected binaries account for $\sim$640 mergers per galaxy in a Hubble time, with another 335 coming from those binaries that merge while still in the cluster.  Over a Hubble time this gives a rate of 10$^{-7}$ per year per galaxy.  These are very conservative estimates for the rate, as the Milky Way is assumed to have about 150 globular clusters, and giant ellipticals can have on the order of $10^3$.  Assuming a value of 300 globular clusters per galaxy and 300 black hole binaries per cluster, there would be an order of magnitude jump in the rate to 10$^{-6}$ per year per galaxy.  We also expect a further increase in rate from the additional mergers produced by black holes that are retained after their first merger.  Galactic binary BH merger rates are estimated at $~10^{-6}$ per year \citep{OShaun05a}.

\begin{figure}[ph]
\centerline{
\rotatebox{0}{\scalebox{0.45}{\includegraphics{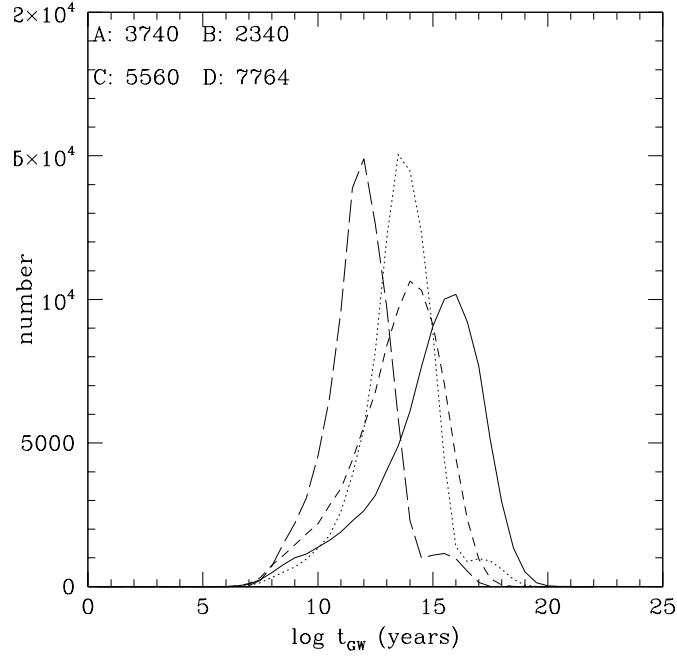}}}
}
\caption{Distributions of $t_{GW}$ for the ejected binaries.  The solid line is for the low mass, low metallicity model (model A), the dotted line is for the high mass, low metallicity model (B), the short dashed line is the low mass, high metallicity model (C), and the long dashed line is the high mass, high metallicity model (D).  We find the effect of both increasing mass and higher metallicity is to shift the distribution to shorter times.  The distributions at lower metallicity are broader, as indicated by the lower peak value.  The peak values for the models are at $10^{16}$, $10^{13.5}$, $10^{14}$, and $10^{12}$ years for models A,B,C, and D respectively.  The number of binaries with $t_{GW}<10^{10}$ is given in the plot.\label{tGW}}
\end{figure}

The mergers are expected to be delayed from the formation of the clusters, which in the case of globulars is close to the beginning of the universe.  The last interaction before the binary is ejected typically happens when the semimajor axis is 0.1-1 AU, giving a $t_{enc}$ of $~10^8-10^9$ years.  The timescales for the gravitational merger of the ejected binaries spans a wide range of values ($5<\log t_{GW}<20$).  Figure \ref{tGW} shows the distribution of merger timescales for ejected binaries for each of the models.  We find that the percentage of binaries which merge between 1 and 10 Gyr is 2.6\% for model A, 2.3\% for model B, 4.3\% for model C, and 8.2\% for model D.  While we have used a thermal distribution ($P(e)=2e$, $\langle e\rangle=0.67$) for the eccentricity after an exchange, the 3-body study by \citet{Sigurdsson93b} found that this works for equal mass exchanges, but for non-equal masses, the eccentricities may be higher ($\langle e\rangle\approx 1-1.3(m_3/m_2)$).  This does not affect recoil velocities, but the $t_{GW}$ would be shortened, and the expected rates of black hole mergers would increase by a factor of a few.  If we choose black hole binaries as we do for the simulation runs and determine their $t_{GW}$ without any interactions, we find that for metal rich systems, only 0.1\% merge in less that a Hubble time, whereas 7\% of low metallicity binaries do so.  This we explain as a model dependent result, the low metallicity period distribution includes systems which have periods shorter than 100 days while the metal rich distribution does not.  Interactions are of great importance in metal rich systems for producing observable mergers, while they are ambivalent in metal poor systems.  

The chirp masses for cluster binary mergers are much higher due to exchanges undergone while the binary was in the cluster.  
We find that, while galactic mergers have chirp masses of 3-8M$_{\sun}$ \citep{Belczynski07}, the chirp masses for the ejected binaries are 15-25M$_{\sun}$ in the metal rich case and 20-40M$_{\sun}$ in metal poor clusters.  The higher chirp masses, while dependent on the models used for the initial mass function of the black holes, is a distinct prediction characteristic of the globular cluster binaries, and easily observable by gravitational radiation instruments.  Since the strain due to gravitational radiation scales as $M/r$, the factor of 4-6 increase in mass of the cluster binaries makes them visible over a factor of 60-200 larger volume, which makes them almost as important source as galactic binaries for the conservative values of GC/galaxy and binaries/GC.  If we assume the less conservative numbers, the cluster binary inspirals would dominate the signal.  LIGO will have an abundance of targets from the ejected binaries.

\begin{figure}[ph]
\centerline{
\rotatebox{0}{\scalebox{0.6}{\includegraphics{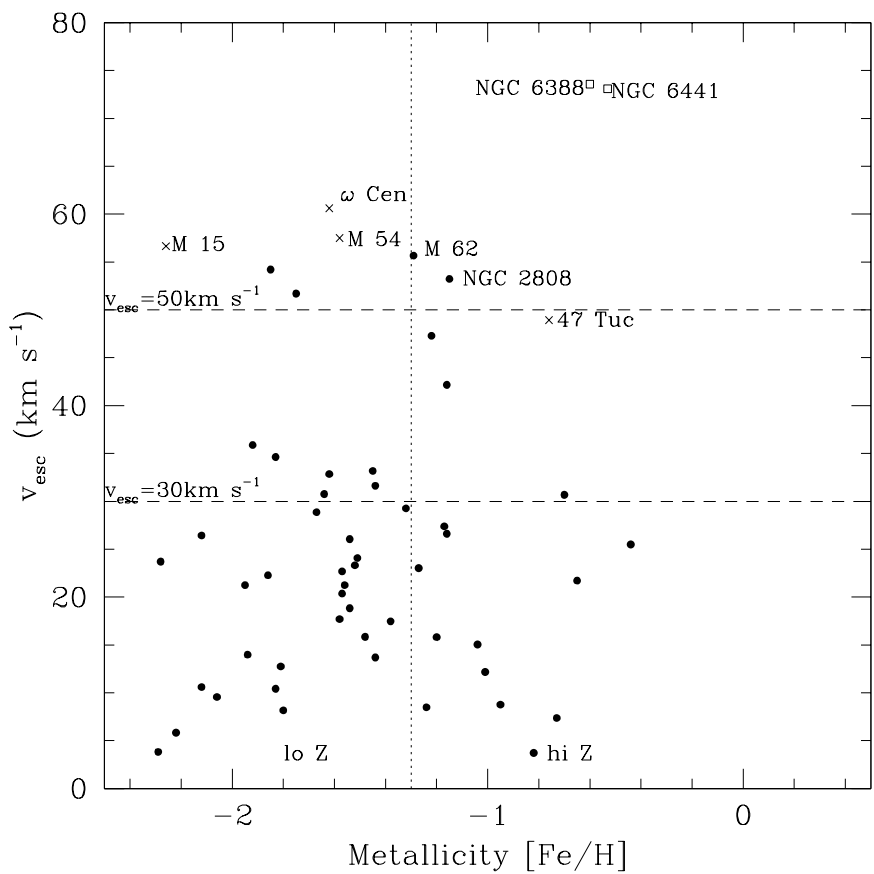}}}
}
\caption{Metallicity versus escape velocity for Milky Way globular clusters.  Horizontal lines show the two escape velocities examined, the vertical line is set at the value of the lower metallicity (Z=0.001).  The two objects in the 
upper right section (open squares) are NGC 6388 and NGC 6441.  Crosses are large well studied globular clusters as labeled.  Suspected dwarf galaxy nuclei are M54/Sgr \citep{Ibata94} and $\omega$ Cen \citet{Norris96,Norris97}.\label{metvesc}}
\end{figure}

Our work in this paper provides a first step from population synthesis to the possible formation of an IMBH in the center of a globular cluster.  Examining a second merger once the merged object has exchanged into a new binary is beyond the scope of this work, but has been studies by \citet{HB07}.  To connect our theoretical models to observed clusters, a plot of metallicity versus $v_{esc}$ is given in figure~\ref{metvesc} using metallicity data from the catalog by \citet{Harris96} and the escape velocity as described above.  Two clusters that have both high metallicity and $v_{esc}>50$ km sec$^{-1}$ are NGC 6388 and NGC 6441.  Both of these clusters lie within 4 kpc of the galactic center.  These clusters are most well known for their contribution to the "second parameter" problem in that they have more extended blue horizontal branches than their metallicity would indicate.  It is speculated that this might be due to dynamical interactions in the clusters \citep{Rich97,Miocchi07}.  They note that the M31 cluster G1, a cluster suspected of having an IMBH by \citet{GRH02}, also shows an extended blue horizontal branch.  These clusters may have been at one point the nuclei of dwarf galaxies, as a couple of other suspected nuclei appear in interesting regions of the metallicity-$v_{esc}$ plot.  Other clusters suspected of being dwarf galaxy nuclei are M54 (due to its association with the Sagittarius dwarf galaxy) by \citet{Ibata94} and $\omega$ Cen \citep{Norris96,Norris97}.  The presence of extended blue horizontal branch stars in the metal-rich clusters NGC 6388 and NGC 6441 is thought to give a similar argument for their being formed in a similar manner \citep{Piotto97}.  A couple others which are less outstanding but still in the upper right part of the diagram are NGC 2808 and M62.  Observations of variability in the recently discovered ULX in a globular cluster of NGC 4472 by \citet{Maccarone07} lead to estimates of a 300M$_{\sun}$ IMBH, though their other solution gives a mass of 30M$_{\sun}$.  They find a metallicity of the cluster of -1.7 from color-metallicity relations, and the luminosity gives it a absolute magnitude of -9.2.  When compared to analogous clusters in the Milky Way (e.g. NGC 6273), this cluster fits into category C, low metallicity high mass.  Figure~\ref{metvesc} places NGC 6388 and NGC 6441 in context with other massive, well studied globular clusters.  

In conclusion, we find that within our simplified model assumptions, most black hole binaries are ejected through gravitational 3-body interaction from the cluster into the general potential of the galaxy.  Of those binaries that survive to merge by gravitational radiation, about 2/3 to half are ejected through gravitational radiation recoil.  Between 0.5\% and 3.5\%, depending on metallicity and cluster escape velocity, of all black hole binaries in clusters are predicted to be retained upon merger of the binary, with typical final masses of 20-50 M$_\sun$, but in some instances over 100M$_{\sun}$.  Of course if other formation channels dominate, or there is significant gas accretion after the dynamical interaction phase, then the final black hole masses may be very different (higher if there is significant accretion).  We find that the rate per galaxy of black hole binary mergers from the globular cluster population, through gravitational radiation is competitive with the total merger rate from the parent galaxy, but biased towards higher masses.  While most globular clusters in massive galaxies probably form at high redshift, this suggests that black hole binary coalescence from clusters in low mass, nearby star forming galaxies may be a significant contributor to the total high frequency gravitational radiation signal in the local universe.  The current results are dependent on the exact form of the initial conditions of mass distributions and period distributions obtained from population synthesis.  As the formation mechanisms for black holes become more well understood, it would be appropriate and easy to refine the results in this paper.

We thank the Center for Gravitational Wave Physics for support, and Ben Owen and Richard O'Shaughnessy for helpful discussions in making this paper.  We would also like to thank the referee for his helpful suggestions.  This research was funded by NSF grant PHY 0114375.

\begin{deluxetable}{cccc}
\tablecolumns{3}
\tablenum{1}
\tablewidth{0pt}
\tablecaption{Initial conditions of the models.\label{modeltab}}
\tablehead{\colhead{Model} & \colhead{metallicity} & \colhead{$v_{esc}$(km/sec)}}
\startdata
A & low & 30\\
B & high & 30\\
C & low & 50\\ 
D & high & 50\\
\enddata
\tablecomments{All runs used $\sigma_{GC}$=10 km/sec and $\Delta$=0.4 as 
described in text.}
\end{deluxetable}

\begin{deluxetable}{ccccc}
\tablecolumns{5}
\tablenum{2}
\tablewidth{0pt}
\tablecaption{Results of simulations.\label{restab}}
\tablehead{\colhead{Model} & \multicolumn{3}{c}{Fate(\%)} & \colhead{\% ejected 
binaries}\\ & 1 & 2 & 3 & with $t_{GW}<t_{H}$ }
\startdata

A & 95.7 (95744) & 3.8 (3790) & 0.5 (466) & 5.3 (5071)\\
B & 98.2 (98242) & 1.0 (997) & 0.8 (761) & 4.1 (4023)\\
C & 94.1 (94069) & 4.5 (4530) & 1.4 (1401) & 8.3 (7799)\\ 
D & 92.7 (92666) & 3.8 (3842) & 3.5 (3492) & 13.3 (12362)\\
\enddata
\tablecomments{Percentages given, absolute number out of $10^5$ in parentheses.  Fate:  1--binary ejected by 3-body interaction, 2--binary ejected upon merger by gravitational radiation recoil, 3--merged binary retained in globular cluster.  Column (5) shows percentage (number) of binaries in Fate 1 with $t_{GW}<10^{10}$ years.}
\end{deluxetable}

\clearpage

\end{document}